%% file: SLR-paper-capita.tex
\documentclass{llncs}
\usepackage{multibib}
\newcites{noslr}{References not Part of the SLR}

\usepackage{geometry}
\usepackage{makeidx}  
\usepackage[utf8]{inputenc}
\usepackage{chngpage}
\usepackage{letltxmacro}
\LetLtxMacro{\oldcite}{\cite}
\renewcommand{\cite}[1]{\mbox{\oldcite{#1}}}

\usepackage{caption}
\renewcommand{\figurename}{Figure}

\usepackage{graphicx}

\usepackage{hyperref}
\usepackage[]{cleveref}

\usepackage{lipsum}

\usepackage{multirow}
\usepackage[table,xcdraw]{xcolor}

\usepackage{dirtytalk}
\usepackage{float}
\usepackage[]{hypcap}
\usepackage[export]{adjustbox}
\usepackage{longtable}
\usepackage{verbatim}
\usepackage{tablefootnote}
\usepackage{tcolorbox}

\usepackage{cite} 

\begin{document}
%

\frontmatter          
\pagestyle{headings}  

\mainmatter

\title{Identification of Practices and Capabilities in API Management: A Systematic Literature Review}

\titlerunning{API}

\author{Max Mathijssen\textsuperscript{a}, Michiel Overeem\textsuperscript{a,}\textsuperscript{b}, Slinger Jansen\textsuperscript{a}}

\institute{\textit{\textsuperscript{a}Utrecht University\\ Department of Information Sciences\\
\{m.mathijssen, m.overeem, slinger.jansen\}@uu.nl\\\ \textsuperscript{b}AFAS Software\\michiel.overeem@afas.nl
}}

\authorrunning{Max Mathijssen}


\maketitle
\medskip
\begin{abstract}
Traditional organizations are increasingly becoming software producing organizations. This software is enabling them to integrate business processes between different departments and with other organizations through Application Programming Interfaces (APIs). The main task of managing APIs is to ensure that the APIs are easy to use by third parties, such as providing helpful documentation, monitoring API performance, and even monetizing API usage. The knowledge on API management is scattered across academic literature. In this document, we describe a Systematic Literature Review (SLR) that has the goal of collecting API Management practices and capabilities related to API Management, as well as proposing a comprehensive definition of the topic. In the scope of this work, a practice is defined as any practice that has the express goal to improve, encourage and manage the usage of APIs. Capabilities are defined as the ability to achieve a  certain  goal  related  to  API  Management,  through  the  execution  of  two  or  more  interrelated practices.We follow a standard method for SLRs in software engineering. We managed to collect 24 unique definitions for the topic, 114 practices and 39 capabilities.
\end{abstract}
\medskip

\clearpage
\input{1Introduction.tex}
\input{2ReviewProtocol.tex}
\input{3LiteratureBody.tex}
\input{4Analysis.tex}
\input{5FutureWork.tex}
\input{6Conclusion.tex}
\input{7Appendices.tex}

\newpage
\bibliographystylenoslr{abbrv}
\bibliographynoslr{References}

\newpage
\renewcommand{\refname}{Publications part of the SLR}
\bibliographystyle{abbrv}
\bibliography{References}

\end{document}

%% file: 1Introduction.tex
\section{Introduction} \label{Section:Intro}
In recent years, there has been an increasing demand among organizations to have access to enterprise data through a multitude of digital devices and channels. In order to meet these expectations, enterprises need to open and provide access to their assets in an agile, flexible, secure and scalable manner \cite{de2017api}. This is accomplished by utilizing Application Programming Interfaces (APIs), which expose an enterprise's data and services to (third party) consumers by allowing applications to easily communicate with one another. However, after an API has been created, it needs to be managed so that developers may easily integrate it into their applications. This is accomplished by performing activities such as providing helpful documentation, controlling and monitoring access to the API, as well as monitoring and analysing its usage. Oftentimes, these activities are supported through an integrated API Management platform, which helps an organization publish APIs to internal, partner, and external developers to unlock the unique potential of their assets \cite{de2017api}.\\ However, despite growing interest in the topic of API Management, more research is needed in order to fill knowledge gaps and identify best practices regarding the subject. This is highlighted by the observation that currently, relatively little literature on API Management exists. As a result, no frameworks or overviews that capture all the practices, capabilities and features API Management is comprised of and a uniform, comprehensive and widely accepted definition of the topic is lacking within the research community.\\ We have conducted a Systematic Literature Review (SLR) in the field of API Management, which is based on the methodology developed by Okoli \citenoslr{okoli2015guide}, as well as guidelines composed by Kitchenham \citenoslr{kitchenham2007guidelines}. The main objectives of this literature review are two-fold. First, we aim to provide a comprehensive overview of literature related to the topic at hand. This is accomplished by first identifying and analyzing the different definitions of the subject, which are then subsequently used as input towards proposing a comprehensive definition of API Management.\\ Secondly, the API Management features are identified and extracted in the form of practices and capabilities, which are components of the Focus Area Maturity model, as first introduced by Steenbergen et al \citenoslr{van2010design}. Maturity models are a proven tool used in the creation of collections of knowledge of practices and processes concerning a particular domain~\citenoslr{jansen2020focus,becker2009developing}. One specific type of maturity model is the Focus Area Maturity model (FAMM) \citenoslr{van2010design,van2013improving}, which is used to establish the maturity levels of an organization in a specific functional domain. This functional domain is described by the set of focus areas that constitute it \citenoslr{jansen2020focus}. Each focus area is composed out of a set of capabilities. These capabilities are positioned against each other in a maturity matrix. In the scope of this research, capabilities are defined as the ability to achieve a  certain  goal  related  to  API  Management,  through  the  execution  of  two  or  more interrelated practices. In this work, the practices that capabilities are composed of are defined as any practice that has the express goal to improve, encourage and manage the usage of APIs. Based on the positioning of the capabilities and practices in the maturity matrix, a number of maturity levels may then be distinguished \citenoslr{jansen2020focus}. These maturity levels may then be used to guide an organization in the incremental development of the functional domain. For example, Spruit \& Röling \citenoslr{spruit2014isfam} have developed a FAMM which may be used by organizations to determine their current information security maturity level, and in recent work by Jansen \citenoslr{jansen2020focus} a FAMM is presented with which organizations can assess their ecosystem governance practices. \\ The rest of this document is structured as follows: in \textbf{Section 2} the review protocol this review is based on is specified. In \textbf{Section 3}, the manner in which the included body of literature is collected, extracted and coded is described. Then, in \textbf{Section 4}, we analyze the collected literature, answering the research questions posed in section 2 in the process. Finally, in \textbf{Section 5} we provide directions on future work which needs to be done and conclude our work in \textbf{Section 6}.

%% file: 2ReviewProtocol.tex
\section{Systematic Literature Review Protocol} \label{Section:2ReviewProtocol}

 The review protocol that was applied as part of the Systematic Literature Review (SLR) which is described throughout this work is based on the methodology developed by Okoli \citenoslr{okoli2015guide}, as well as guidelines composed by Kitchenham \citenoslr{kitchenham2007guidelines}. Consisting of several steps which should be adhered to, Okoli's methodology is aimed towards assisting researchers in carrying out a standalone SLR, while Kitchenham \citenoslr{kitchenham2007guidelines} presents general guidelines for undertaking systematic reviews. Adhering to these will ensure that this literature review is  explicit in explaining the procedures by which it was conducted, comprehensive  in  its  scope  of  including  all  relevant  material, and hence,  reproducible  by  others  who  wish  to  follow  the  same  approach  in reviewing  the topic  of  API  Management.
 
 \subsection{Research Questions}
The establishment of the review protocol described in this section is necessary to ensure that the literature review is systematic, minimizing researcher bias as a result. In order to accomplish this, the literature review is guided by two research questions that serve the aim of this work and highlight the motivation that initiated this review. The following research questions will be addressed:\\

\begin{adjustwidth}{0.5cm}{}\textbf{RQ1}: \textit{How is the topic of API Management characterized?} \end{adjustwidth}
\begin{adjustwidth}{1.5cm}{}This research question is addressed by answering the two sub research questions (SubRQ1.1 and SubRQ1.2) listed below. First, definitions of API Management are extracted as a result of the Systematic Literature Review, after which these are analyzed in order to form a basis for formulating a comprehensive definition of the topic at hand. \\ \end{adjustwidth}

\begin{adjustwidth}{1cm}{} \textbf{SubRQ1.1}: \textit{How is API Management defined and interpreted in academia?}\end{adjustwidth}\ \begin{adjustwidth}{3cm}{}In order to properly analyze the field of API Management, this term should first be defined as an object of study. As such, the first objective of this work is to provide an overview of how the term 'API Management' is defined within the research community. This is achieved by examining and comparing the various definitions encountered within the body of relevant literature, which is produced as a result of the Systematic Literature Review. \\ \end{adjustwidth}

\begin{adjustwidth}{1cm}{} \textbf{SubRQ1.2}: \textit{What comprehensive definition for API Management may be constructed based on the findings of a Systematic Literature Review?} \end{adjustwidth} \begin{adjustwidth}{3cm}{} Given the fact that API Management is a relatively new concept, there is no universal and comprehensive definition in use across literature on the subject. Instead, a plethora of definitions which focus on varying aspects and perspectives of API Management can be found within the research community. Because of this, an all-encompassing and comprehensive definition should be developed. \\ \end{adjustwidth}

\begin{adjustwidth}{0.5cm}{}\textbf{RQ2}: \textit{What practices and capabilities may be identified and extracted from the existing body of literature related to API Management?} \end{adjustwidth}
\begin{adjustwidth}{1.5cm}{} After having defined the subject of API Management, the main objective of this work is shifted towards the identification and extraction of practices and capabilities as encountered in the literature related to API Management. These concepts are components of the Focus Area Maturity Model, as described and defined in sections 1 and 4.2 of this work.
\end{adjustwidth}

 \subsection{Defining the Literature Body}
 The strategy employed for collecting relevant literature first commences by performing a keyword search in a selected list of scientific libraries. The scientific libraries included in this search are:
 
 \begin{enumerate}
     \item The ACM Digital Library \footnote{https://dl.acm.org/}
     \item DBLP: Computer Science Bibliography \footnote{https://dblp.uni-trier.de/}
     \item IEEE Computer Society Digital Library \footnote{https://www.computer.org/csdl/home}
     \item Google Scholar \footnote{https://scholar.google.com/}
     \item Scopus \footnote{https://www.scopus.com/}
 \end{enumerate}
 
 The initial extraction of literature consists of a keyword search, with the search terms "API" AND ("management" OR "gateway" OR "strategy" OR "practice" OR "capability" OR "lifecycle" OR "versioning") being entered in each of the libraries listed above. 
 
 This search query was composed as based on an initial exploration of the topic of API management. This consisted of discussions among the authors of this work, a superficial literature search and several exploratory interviews which were conducted with API architects in the software development industry. The 'gateway' keyword was included based on the initial observation that in a significant amount of papers, the API gateway is either used as a synonym for API management or considered to be an integral component of API management. In the same vein, the keyword 'strategy' was included based on the hypothesis that enterprises may regard API management as a strategy which may be utilised to improve business processes. Furthermore, the 'practice' and 'capability' keywords were added to the search query in hopes of identifying best practices and features as part of API management. Lastly, the 'lifecycle' and 'versioning' keywords were included based on the observation that these concepts are closely interrelated to the topic of API management. The resulting search query is broad in terms of its scope. This choice was made deliberately and intentionally, mainly due to the fact that relevant literature related to the topic at hand is scarce and scattered across publications of varying disciplines and domains. \\ \\ As a result, executing the search query results in an initial body containing the maximum amount of relevant articles with regards to the scope of this research. Next, the selected literature body collected from the aforementioned scientific libraries should commit to a set of inclusion criteria:

 \begin{itemize}
     \item Is written in English.
     \item Has a document body consisting of more than one page.
     \item Is a book, research paper, thesis or whitepaper.
     \item Is publicly accessible.
     \item Is published in or after the year 2010.
     \item The literature should address API Management as an area of research, either primarily or secondarily. This holds that the keywords "API" AND ("management" OR "gateway" OR "strategy" OR "practice" OR "capability" OR "lifecycle" OR "versioning") should be encountered in either the title, keywords section or abstract.
 \end{itemize}
 
 As a result, the initial body of selected literature does not contain document types such as extended abstracts, presentations, keynotes or books and papers written in other languages than English. The final included body of literature was formed through the execution of the following steps:
 
 \begin{enumerate}
     \item \textit{Collecting all initial literature.} The intial literature collection is the result of the scientific library portal search. This search is executed by querying the complete text body, abstract, title and keyword sections using the earlier mentioned keywords "API" AND ("management" OR "gateway" OR "strategy" OR "practice" OR "capability" OR "lifecycle" OR "versioning"), so that the maximum amount of possibly relevant literature is included.
     \item \textit{Applying inclusion criteria}. The body of literature which was produced as a result of the previous step is searched for the presence of the aforementioned keywords in the title, abstract and keyword sections. Additionally, the set of inclusion criteria which were defined above are applied.
     \item \textit{Removing duplicates}. The reference lists corresponding to each scientific library, containing literature which passed the application of the inclusion/exclusion criteria as part of the previous step, were cross-referenced with one another. This was deemed necessary largely due to the fact that the Google Scholar database was included in the literature collection process and that this portal aggregates publications from the other selected databases used in this review. After having cross-referenced all literature that was selected for inclusion at this step of the collection process, any duplicate literature was removed.
     \item \textit{Applying exclusion criteria}. Next, all books, research papers, theses and whitepapers contained in the remaining collection of literature were scanned for the presence of any definitions, practices or capabilities. Literature in which none of these elements appear were excluded. In order to properly match the scope of our research, the given definition for practices as defined in section 2.1 of this work is modified into the following: a practice is defined as any practice that has the express goal to improve, encourage and manage the usage of APIs. 
 \end{enumerate}

In addition to being based on the methodology developed by Okoli \citenoslr{okoli2015guide} and guidelines composed by Kitchenham \citenoslr{kitchenham2007guidelines}, the stepwise SLR protocol described above is inspired by Manikas \& Hansen's \citenoslr{manikas2013software} literature review of Software Ecosystems. 

%% file: 3LiteratureBody.tex
\section{Collecting the Literature Body} \label{Section:3LiteratureBody}

In order to obtain the initial literate body serving as input for this review, the systematic literature review (SLR) protocol described in the previous section of this work is carried out on the extraction date of April 24, 2020. As an overview, the four steps which are executed in order to define the definitive literature body for this review along with the number of papers they produced as a result can be seen in \textbf{Table~\ref{tab:SLRSteps}} below. 

\setlength{\tabcolsep}{10pt}
\renewcommand{\arraystretch}{1.2}
\begin{table}[h!]
\centering
 \begin{tabular}{||l r||} 
 \hline
\textbf{Step} &\textbf{ Number of papers} \\ [0.5ex] 
 \hline\hline
 1. Collecting all initial literature  &  5152 \\ 
 2. Applying inclusion criteria &  117 \\
 3. Removing duplicates &  78 \\
 4. Applying exclusion criteria & 43 \\[1ex] 
 \hline
 \end{tabular}
 \caption{\label{tab:SLRSteps}Steps and corresponding number of papers as part of the literature collection process.}
\end{table}

Initially, the literature collection consisted of 5132 papers which were extracted from the five libraries, using the keywords as described in section 2. However, it should be noted that this number is highly inflated due to the inclusion of the Google  Scholar  database  as part of the  literature  collection  process, considering that this portal aggregates publications from the other selected databases used in this review. An overview of this initial literature collection procedure can be seen in \textbf{Table~\ref{tab:NoOfPapersPerDB}} below, showing the number of papers that were identified as grouped by each individual scientific library and search term.\\

\renewcommand{\arraystretch}{1.2}
\begin{table}[h!]
\footnotesize
\centering
\setlength\tabcolsep{6pt}
 \begin{tabular}{||l|rrrrrrr|r||} 
 \hline
\rotatebox[origin=c]{270}{Database} & \rotatebox[origin=c]{270}{Management} & \rotatebox[origin=c]{270}{Gateway} & \rotatebox[origin=c]{270}{Strategy} & \rotatebox[origin=c]{270}{Practice} & \rotatebox[origin=c]{270}{Capability} & \rotatebox[origin=c]{270}{Lifecycle} & \rotatebox[origin=c]{270}{Versioning} & \rotatebox[origin=c]{270}{Total} \\ [0.5ex] 
 \hline\hline
 ACM & 98 & 95 & 11 & 1 & 18 & 8 & 19 & 250\\ 
 DBLP & 82 & 5 & 8 & 18 & 4 & 2 & 2 & 121\\
 IEEE & 5 & 17 & 0 & 0 & 1 & 0 & 0 & 23\\
 Google Scholar & 1630 & 2040 & 248 & 78 & 100 & 108 & 240 & 4444\\
 Scopus & 114 & 116 & 23 & 28 & 22 & 5 & 6 & 314\\ \hline
 \textbf{Total} & \textbf{1929} & \textbf{2273} & \textbf{290} & \textbf{125} & \textbf{145} & \textbf{123} & \textbf{267} & \textbf{5152} \\ [1ex] 
 \hline
 \end{tabular}
 \caption{\label{tab:NoOfPapersPerDB}Number of papers identified as grouped by database and search term.}
\end{table}

Next, after having applied the set of inclusion criteria mentioned in section 2, 5025 papers are rejected. As a result, 117 papers are included in the resulting literature body after this step. This body of literature was then imported into a central database using Nvivo 12 Pro, which is accessible to all authors and is publicly available on Mendeley \footnote{\url{https://data.mendeley.com/datasets/3k3fjrkbnj/draft?a=a910cb61-5412-4865-8496-34775bec5563}}. The manner in which this database was created, structured and may be used by anyone wishing to do so is described below: \\

\begin{adjustwidth}{0.5cm}{}
The Nvivo database's data repository consists of four folders. Literature which has passed all inclusion and exclusion criteria is located in the 'Accepted' folder and contains at least one practice or capability, and optionally, a definition. Literature in which a definition for the topic of API Management was found but no practices or capabilities, is located in the 'Definitions' folder. Literature which has failed to pass all inclusion and exclusion criteria is located in the 'Rejected' folder. The 'Figures' folder contains any relevant figures that were encountered in the body of collected literature with regards to the scope of the literature review.
\end{adjustwidth}
\renewcommand{\arraystretch}{1.2}
\begin{table}[h!]
\footnotesize
\centering
 \begin{tabular}{||l| l l l l l l l|r||} 
 \hline
\rotatebox[origin=c]{270}{Database} & \rotatebox[origin=c]{270}{Management} & \rotatebox[origin=c]{270}{Gateway} & \rotatebox[origin=c]{270}{Strategy} & \rotatebox[origin=c]{270}{Practice} & \rotatebox[origin=c]{270}{Capability} & \rotatebox[origin=c]{270}{Lifecycle} & \rotatebox[origin=c]{270}{Versioning} & \rotatebox[origin=c]{270}{Total} \\ [0.5ex] 
 \hline\hline
 ACM & 13 & 0 & 1 & 0 & 0 & 0 & 2 & 16\\ 
 DBLP & 10 & 0 & 1 & 2 & 0 & 0 & 0 & 13\\
 IEEE & 4 & 0 & 0 & 0 & 0 & 0 & 0 & 4\\
 Google Scholar & 43 & 7 & 3 & 0 & 0 & 4 & 2 & 59\\
 Scopus & 14 & 5 & 1 & 1 & 0 & 1 & 3 & 25\\ \hline 
 \textbf{Total} & \textbf{84} & \textbf{12} & \textbf{6} & \textbf{3} & \textbf{0} & \textbf{5} & \textbf{7} & \textbf{117} \\ [1ex] 
 \hline
 \end{tabular}
 \caption{\label{tab:AllPapersPerDatabase}Number of papers adhering to inclusion criteria as grouped by database and search term.}
\end{table}
\begin{adjustwidth}{0.5cm}{}
The database's node repository consists of four folders. The 'Capabilities' folder contains all coded capabilities with relation to API Management, which were encountered during the scanning of the collected body of literature. Upon the first and initial observation of a capability, a new node was created and named within this folder. Any relevant text describing the newly discovered capability was then highlighted and coded to this node. Subsequent discoveries of identical capabilities across other bodies of work were then highlighted and coded to the preexisting node. Similarly, this procedure was executed for all nodes contained in the 'Practices', 'Definitions' and 'Figures' folders.\\ \end{adjustwidth}

Roughly 10\% of bodies of literature were selected at random, and were subsequently checked for inter-rater agreement among all authors. In the event of the coding of any practice, capability or definition that one of the authors did not agree upon, any discrepancy was discussed and corrected as part of a monthly meeting with all authors. As a result of performing this inter-agreement check, codings were then either left unaltered, edited or removed. In doing so, the construct validity of codings is ensured \citenoslr{eisenhardt1989building}.\\
An overview of the number of papers adhering to the inclusion criteria as grouped by database and search terms can be seen in \textbf{Table~\ref{tab:AllPapersPerDatabase}}. As part of the third step, the reference lists corresponding to each scientific library, which contain literature that passed the application of the inclusion/exclusion criteria as part of the previous step, were cross-referenced with one another. Doing so resulted in the removal of 39 duplicate papers. As part of the fourth step, all remaining books, research papers, theses and white papers were manually scanned for the presence of any definitions, practices, capabilities or focus areas regarding the subject of API management. Literature which does not include any of these elements was excluded from this review. This was the case for 35 papers, resulting in a final body of literature of 43 papers. Out of this final collection of literature, 11 papers contained a definition of API Management but no corresponding practices or capabilities. Furthermore, 16 papers contained at least one practice or capability but no definition, while 16 papers contained both a definition as well as at least one practice and/or capability. 

%% file: 4Analysis.tex
\setlength{\tabcolsep}{10pt}
\renewcommand{\arraystretch}{1.2}
\begin{table}[h!]
\centering
 \begin{tabular}{||l l r||} 
 \hline
\textbf{Year} & \textbf{Papers} & \textbf{Total} \\ [0.5ex] 
 \hline\hline
 2011  & ~\cite{jacobson2011apis,raivio2011towards} & 2  \\ 
 2012  &  None. & 0 \\
 2013  & ~\cite{thielens2013apis} & 1 \\
 2014  & ~\cite{krintz2014cloud,o2014internet,hofman2014technical} & 3 \\
 2015  & ~\cite{biehl2015api,familiar2015iot,fremantle2015web,gamez2015towards,jayathilaka2015eager,sine2015api,weir2015oracle} & 7 \\ [1ex] 
 2016  & ~\cite{liang2016exploiting,montesi2016circuit,sutherland2016investigation} & 3 \\ [1ex] 
 2017  & ~\cite{ciavotta2017microservice,de2017api,matsumoto2017fujitsu,mussig2017highly,nakamura2017fujitsu,namihira2017iot,patni2017pro,Yu2017}  & 8 \\ [1ex] 
 2018  & ~\cite{gadge2018microservice,haselbock2018microservice,indrasiri2018developing,medjaoui2018continuous,preibisch2018api,vsnuderl2018rate,vijayakumar2018practical,Zhao_2018} & 8 \\ [1ex] 
 2019  & ~\cite{akbulut2019software,ala2019application,andreo2019api,coste2019api,hamalainen2019api,hohenstein2018architectural,andrey_kolychev_konstantin_zaytsev_2019_3256462,lourencco2019framework,santana2019case,Xu_2019} & 10 \\ [1ex] 
 \hline
 \end{tabular}
 \caption{\label{tab:YearOfPubTable}Papers categorized by their year of publication.}
\end{table}

\section{Analysis} \label{Section:4Analysis}
In this section of this work,the final body of literature and the results of the review are analyzed. As mentioned earlier ,this final collection of literature consists of 43 books,research papers, theses and white papers. The year in which these were published range from the year 2011 to 2019. Publications stemming from the year 2010 and earlier were excluded from the literature collection procedure due to the fact that API Management is a topic that has emerged in recent years. The collected literature on the subject of API Management is ordered according to their publication year, as can be seen in \textbf{Table \ref{tab:YearOfPubTable}} below. The included literature on API Management originating from the years 2011, 2012 and 2013 is scarce. However, from the year 2014 onwards, a noticeable surge in the amount of published papers emerges. While this observation may be skewed due to the inclusion and exclusion criteria used in this literature review, this recent rise in the amount of publications on API Management may signal an increase in the importance of and interest in the subject among the research community.

For the following analysis of the final included body of literature, an overview of the definitions which were extracted from the body of literature is first presented and analyzed. Then, a key word frequency analysis is performed on these definitions, which is then used as input for the formulation and proposition of an all-encompassing and comprehensive definition for the topic of API Management. Lastly, the collection of capabilities and practices which were extracted from the literature is presented and analyzed. As a result, the research questions posed in section 2.1 are answered. 

\subsection{Defining API Management}
During this literature review, an overview of definitions for the topic of API Management was collected. This was done by scanning and coding the 78 books, research papers, theses and white papers that were produced as a result of the collection procedure as described in section 3 of this work. Among the final body of included literature, which consists of 43 papers, 27 papers contained a definition for API Management. In this collection of 27 papers, 24 unique definitions for API Management were identified. An complete overview of all 24 of these definitions may be reviewed in \textbf{Appendix A}. \\ 

Out of the 43 included papers, 16 papers did not contain any definition for API Management but did contain at least one practice and/or capability related to the topic. In explaining this observation, three main reasons for the lack of a definition of the topic at hand were identified.\\ For the largest portion of these papers, the reason as to why a definition is missing is that while a (architectural) component related to API Management is discussed, the topic as a whole is out of the scope of the research which is conducted. For example, Abkulut \& Perros \cite{akbulut2019software} present a version management approach utilizing the API gateway design pattern. Despite the fact that the API gateway is considered to be an integral architectural component of most API Management solutions, the broader topic of API Management is out of the scope of Abkulut \& Perros' research. As a result, this paper contains practices and/or capabilities but no definition of the subject, due to the fact that the approaches and techniques related to versioning, gateways and scaling discussed in this work were interpreted as practices and capabilities in the scope of API management by the authors of this SLR. Similarly, this is the case for research performed by Ciavotta et al. \cite{ciavotta2017microservice}, Gadge \& Kotwani \cite{gadge2018microservice}, Gamez Diaz et al. \cite{gamez2015towards}, Montesi \& Weber \cite{montesi2016circuit}, Müssig et al. \cite{mussig2017highly}, Preibisch \cite{preibisch2018api}, Xu et al. \cite{Xu_2019} and Zhao et al. \cite{Zhao_2018}. \\ The second group of papers whose authors failed to include a definition for API management consists of those which simply do not mention it. For example, while Jacobson's book on API strategy \cite{jacobson2011apis} does not discuss API Management as a central topic, this work does in fact devote a separate chapter on the topic. Even though this chapter does contain practices and capabilities related to API management and the need for managing APIs is highlighted, a clear definition is missing. Similarly, in Raivio et al's \cite{raivio2011towards} paper on API management providers where API management may be considered to be the main topic under investigation, a definition is not given. Other papers where a definition is missing despite the fact that API management is one of the main topics discussed, are those of Šnuderl \cite{vsnuderl2018rate} and Hofman \& Rajagopal \cite{hofman2014technical}.\\
The last group of literature in which no definition for API Management was found, consists of papers which focus on a domain that is closely related to API management or may be regarded to be within the scope of this topic. For example, in their work, Jayathilaka et al. \cite{jayathilaka2015eager} investigate a new approach to API governance, which the authors of this SLR have identified to be a subtopic within the field of API management. Additionally, Jayathilka et al's paper contains practices and capabilities related to the API gateway and versioning. However, due to the fact that API management is not the primary topic discussed, a definition for it is missing. Similarly, the same obversation holds true for Krintz et al's work \cite{krintz2014cloud}, in which API governance is also considered to be the main topic under investigation.

Out of the 43 included papers, 27 papers contain a definition for API Management. Interestingly, among this collection of literature, the authors of 21 of these papers include their own uniquely formulated definition of the topic. O'Neil \cite{o2014internet} presents the oldest definition of API Management among the included literature, stemming from 2014:\\

\begin{adjustwidth}{0.5cm}{} \textit{"The API management solution addresses privacy and security issues through monitoring and visibility capabilities, as well as providing audit trails detailing how its APIs are being used. Essentially, the API management solution will act as a gatekeeper, providing  a  reliable  pathway to control and oversee the flow data between the different connected devices."}\\ \end{adjustwidth} 

Furthermore, some authors cite other authors' definitions. For example, Andreo \& Bosch \cite{andreo2019api} cite a definition given in a blog post by Stafford \footnote{\url{https:// searchmicroservices.techtarget.com/feature/Why-use-new-lifecycle-tools-inAPI-management-platforms}}:\\
\begin{adjustwidth}{0.5cm}{} \textit{"In this paper we consider the term API management as described by  Jan   Stafford where he compares, the API  management to application  life-cycle management: observing and  controlling  an API from creation to retirement."}\\ \end{adjustwidth} 

\noindent Hohenstein et al. \cite{hohenstein2018architectural} refer to Wikipedia  \footnote{\url{https://en.wikipedia.org/wiki/API_management}} in defining the subject:\\
\begin{adjustwidth}{0.5cm}{} \textit{"Wikipedia defines API Management as “the process of creating and publishing Web APIs, enforcing their usage policies, controlling  access, nurturing the subscriber  community, collecting and analyzing usage statistics, and reporting on performance.”"} \\\end{adjustwidth}

The 3 remaining authors \cite{ala2019application,buidesign,santana2019case} among the group of those whom do not provide a definition in their own words adopt a definition which is presented by De \cite{de2017api}. Interestingly, De's definition is the only definition which is cited more than once across other papers among the included body of literature. In his book titled \textit{"API Management"}, De interprets API Management as a platform, defining it as follows:\\

\begin{adjustwidth}{0.5cm}{} \textit{"An API management platform helps an organization publish APIs to internal, partner, and external developers to unlock the unique potential of their assets. It provides the core capabilities to ensure a successful API program through developer engagement, business insights, analytics, security, and protection."}\\ \end{adjustwidth}

As an overview, all included literature is presented in \textbf{Table 5} below, accompanied by their corresponding groupings in terms of their absence or presence of definitions for the topic of API Management.

\setlength{\tabcolsep}{10pt}
\renewcommand{\arraystretch}{1.2}
\begin{table}[h!]
\centering
 \begin{tabular}{||l l r||} 
 \hline
\textbf{Definition} &\textbf{Papers} &\textbf{Total} \\ [0.5ex] 
 \hline\hline
 Own  & \cite{biehl2015api,coste2019api,familiar2015iot,fremantle2015web,gamez2015towards,hamalainen2019api,haselbock2018microservice,indrasiri2018developing,andrey_kolychev_konstantin_zaytsev_2019_3256462,liang2016exploiting,lourencco2019framework,matsumoto2017fujitsu,medjaoui2018continuous,nakamura2017fujitsu,namihira2017iot,o2014internet,patni2017pro,sine2015api,sutherland2016investigation,vijayakumar2018practical,weir2015oracle} & 21 \\ 
 None & \cite{akbulut2019software,ciavotta2017microservice,gadge2018microservice,gamez2015towards,jacobson2011apis,jayathilaka2015eager,krintz2014cloud,montesi2016circuit,mussig2017highly,preibisch2018api,raivio2011towards,vsnuderl2018rate,thielens2013apis,hofman2014technical,Xu_2019,Zhao_2018} &  16 \\
 De & \cite{ala2019application,de2017api,buidesign,santana2019case} & 4 \\
 External Source & \cite{andreo2019api,hohenstein2018architectural} & 2 \\[1ex] 
 \hline
 \end{tabular}
 \caption{\label{tab:table-name3}Papers as grouped by their definitions for API Management.}
\end{table}

Now that all definitions for API Management within the body of included literature have been identified and extracted, we are able to formulate a comprehensive definition of the topic. In order to do so, we first perform a key term frequency analysis on the collection of identified definitions. An overview of the results of this analysis may be reviewed in \textbf{Table 6} below.

\setlength{\tabcolsep}{10pt}
\renewcommand{\arraystretch}{1.2}
\begin{table}[h!]
\centering
 \begin{tabular}{||l r||} 
 \hline
\textbf{Word} &\textbf{Frequency} \\ [0.5ex] 
 \hline\hline
Publish & 7 \\
Developer & 7 \\
Control & 7 \\
Platform & 6 \\
Security & 6 \\
Analytics & 6 \\
Gateway & 6 \\
Monitoring & 6 \\
Design & 5 \\
Lifecycle & 4 \\
Consumers & 4 \\
Documentation & 4 \\
Access & 4 \\
Organization & 4 \\
Throttling & 3 \\
Deployment & 3 \\
 \hline
 \end{tabular}
 \caption{\label{tab:table-name}Frequency of key terms encountered across all definitions of API Management.}
\end{table}

When examining the frequency of key terms as encountered across all definitions of API Management, it becomes clear which concepts are deemed to be the most relevant and important towards characterizing the topic at hand. First, it appears that the main actors that are related to performing API Management are the (API) developer and organization, as well as (API) consumers. Secondly, it is found that the main (architectural) component through which API Management is enabled, is the gateway. Thirdly, the topic of API Management often seems to be envisioned as a platform. Lastly, the most frequent capabilities related to API management which are mentioned across all definitions appear to be:

\begin{itemize}
\item Publishing APIs \textbf{(7)}
\item Control of access, authentication, dataflows and the API lifecycle \textbf{(7)}
\item Providing security \textbf{(6)}
\item Providing analytics \textbf{(6)}
\item Providing monitoring \textbf{(6)}
\item Providing tools or support for the design of APIs \textbf{(5)}
\item Providing documentation for APIs \textbf{(4)}
\item Providing usage or rate throttling \textbf{(3)}
\item Enabling the deployment of APIs \textbf{(3)\\}
\end{itemize}

By incorporating these observations, we are able to propose a new and comprehensive definition of API Management. This definition, which contains all of the the key terms that are listed in table 6, is as follows:\\

\begin{adjustwidth}{0.5cm}{} \begin{tcolorbox} \textit{API Management is an activity that enables organizations to design, publish and deploy their APIs for (external) developers to consume. API Management capabilities such as controlling API lifecycles, access and authentication to APIs, monitoring, throttling and analyzing API usage, as well as providing security and documentation are often implemented through an integrated platform, which is supported by an API gateway.} \end{tcolorbox} \end{adjustwidth}

In order to further support the definition given above, we define the keyword 'platform' as follows by adopting the definition provided by De \cite{de2017api} as mentioned earlier in this section:\\
\begin{adjustwidth}{0.5cm}{} \textit{"An API management platform helps an organization publish APIs to internal, partner, and external developers to unlock the unique potential of their assets. It provides the core capabilities to ensure a successful API program through developer engagement, business insights, analytics, security, and protection."}\\ \end{adjustwidth}

\noindent In defining the keyword 'API gateway', we adopt the following definition which is also provided by De \cite{de2017api}:\\ 
\begin{adjustwidth}{0.5cm}{} \textit{"An API gateway forms the heart of any API management solution that enables secure, flexible, and reliable communication between the back-end services and digital apps. It helps to expose, secure, and manage back-end data and services as RESTful APIs. It provides a framework to create a facade in front of the back-end services. This facade intercepts the API requests to enforce security, validate data, transform messages, throttle traffic, and finally route it to the back-end service."} \end{adjustwidth}

\subsection{Identification \& Extraction of practices, capabilities and focus areas}
During  this systematic literature  review, a set of practices and capabilities related to API Management was collected. This was done by scanning and coding the 78 books, research papers, theses and whitepapers that were produced as a result of the collection procedure as described in section 3 of this work. Among  the  final  body  of  included  literature,  which  consists  of  43  papers, 32 papers contained at least one practice or capability.\\ 
In the scope of this research, a practice is defined as follows:\\

\begin{adjustwidth}{0.5cm}{} \begin{tcolorbox} \textit{In the scope of API Management, a practice is any practice that has the express goal to improve, encourage and manage the usage of APIs.} \end{tcolorbox} \end{adjustwidth}

\noindent Furthermore, capabilities are defined as follows:\\

\begin{adjustwidth}{0.5cm}{} \begin{tcolorbox} \textit{A capability is the ability to achieve a certain goal related to API Management, through the execution of two or more interrelated practices.} \end{tcolorbox} \end{adjustwidth}

In total, 114 practices and 39 capabilities were identified by performing the scanning and coding procedure, as described in section 3 of this work. A complete overview of all the practices which were extracted from the included body of literature may be reviewed in \textbf{Appendix B}, and an overview of all extracted capabilities may be reviewed in \textbf{Appendix C}. The overviews presented in these appendices include the names of the practices and capabilities, accompanied by their respective description as well as the respective sources in which they were encountered. In the case of multiple occurrences of a practice or capability across the included literature and the presence of more than one piece of text describing it, the most elaborate and detailed description was selected. In the event where a suitable description or definition of a practice or capability was not able to be identified, the authors provided their own description, which are denoted by italics. Consequently, these descriptions do not originate from literature and are thus not academically grounded. In order to highlight the practices and capabilities that occur most frequently across the studied literature, frequency distributions are presented in \textbf{Table 7} and \textbf{Table 8} below.

\setlength{\tabcolsep}{10pt}
\renewcommand{\arraystretch}{1.2}
\begin{table}[h!]
\centering
 \begin{tabular}{||l l r||} 
 \hline
\textbf{Capability} &\textbf{Frequency} & \textbf{Sources}\\ [0.5ex] 
 \hline\hline
Authentication & 11 & \cite{de2017api,fremantle2015web,gadge2018microservice,gamez2015towards,andrey_kolychev_konstantin_zaytsev_2019_3256462,matsumoto2017fujitsu,montesi2016circuit,vsnuderl2018rate,thielens2013apis,hofman2014technical,Xu_2019}\\
Monitoring & 9 & \cite{biehl2015api,de2017api,gadge2018microservice,jacobson2011apis,medjaoui2018continuous,montesi2016circuit,thielens2013apis,hofman2014technical,Zhao_2018} \\
Security & 9 & \cite{de2017api,gadge2018microservice,jayathilaka2015eager,montesi2016circuit,medjaoui2018continuous,patni2017pro,vijayakumar2018practical,hofman2014technical,Xu_2019} \\
Analytics & 8 & \cite{biehl2015api,de2017api,familiar2015iot,indrasiri2018developing,patni2017pro,vijayakumar2018practical,weir2015oracle,hofman2014technical} \\
Catalog \& Documentation & 8 & \cite{biehl2015api,de2017api,gamez2015towards,jacobson2011apis,lourencco2019framework,patni2017pro,vijayakumar2018practical,hofman2014technical} \\
API Publication \& Deployment & 7 & \cite{biehl2015api,de2017api,lourencco2019framework,medjaoui2018continuous,patni2017pro,vijayakumar2018practical,Xu_2019} \\
Monetization & 6 & \cite{biehl2015api,de2017api,indrasiri2018developing,jacobson2011apis,patni2017pro,weir2015oracle} \\
Version Management & 6 & \cite{akbulut2019software,biehl2015api,de2017api,medjaoui2018continuous,gamez2015towards,vijayakumar2018practical}\\
 \hline
 \end{tabular}
 \caption{\label{tab:table-name1}Frequency of capabilities related to API Management.}
\end{table}

\setlength{\tabcolsep}{10pt}
\renewcommand{\arraystretch}{1.2}
\begin{table}[h!]
\centering
 \begin{tabular}{||l l r||} 
 \hline
\textbf{Practice} &\textbf{Frequency} & \textbf{Sources}\\ [0.5ex] 
 \hline\hline
Caching & 11 & \cite{biehl2015api,de2017api,gadge2018microservice,gamez2015towards,indrasiri2018developing,patni2017pro,preibisch2018api,vsnuderl2018rate,vijayakumar2018practical,hofman2014technical,Zhao_2018}\\
OAuth Authentication & 10 & \cite{de2017api,gadge2018microservice,gamez2015towards,hohenstein2018architectural,matsumoto2017fujitsu,patni2017pro,thielens2013apis,hofman2014technical,Xu_2019,Zhao_2018} \\
Load Balancing & 9 & \cite{biehl2015api,ciavotta2017microservice,de2017api,gadge2018microservice,gamez2015towards,montesi2016circuit,nakamura2017fujitsu,Xu_2019,Zhao_2018} \\
Rate/Quota Limiting & 9 & \cite{de2017api,gamez2015towards,jacobson2011apis,lourencco2019framework,raivio2011towards,jayathilaka2015eager,vsnuderl2018rate,hofman2014technical,gadge2018microservice} \\
Usage Throttling & 9 & \cite{de2017api,fremantle2015web,familiar2015iot,gadge2018microservice,hohenstein2018architectural,indrasiri2018developing,jacobson2011apis,thielens2013apis,weir2015oracle} \\
Activity Logging/Monitoring & 8 & \cite{biehl2015api,de2017api,jacobson2011apis,fremantle2015web,lourencco2019framework,preibisch2018api,sine2015api,thielens2013apis} \\
Access Control & 6 & \cite{de2017api,gadge2018microservice,nakamura2017fujitsu,raivio2011towards,sine2015api,vijayakumar2018practical} \\
Billing & 5 & \cite{de2017api,nakamura2017fujitsu,raivio2011towards,sine2015api,hofman2014technical}\\
 \hline
 \end{tabular}
 \caption{\label{tab:table-name2}Frequency of practices related to API Management.}
\end{table}

\pagebreak

%% file: 5FutureWork.tex
\section{Future Work} \label{Section:5FutureWork}
Based on the results of the systematic literature review presented in the previous sections, it is evident that the research area of API Management is interpreted in a relatively large number of ways within the research community and consists of a plethora of practices and capabilities. Considering the fact that practices and capabilities are the building blocks the Focus Area Maturity Model consists of, these may be used to ultimately construct such a model with which organizations may assess their degree of maturity with regards to API management. First, however, much work remains to be done.\\ As mentioned in previous sections, capabilities describe a goal which may be achieved through the execution of two or more interrelated practices. However, the practices and capabilities related to API Management that were identified in this work are currently not linked to one another. In order to resolve this, any relationships existing between these elements must first be identified, after which practices may be assigned to the capabilities they are related to. In the same vein, capabilities may then subsequently be composed and categorized into focus areas, which are the largest building blocks the Focus Area Maturity Model consists of. Finally, the resulting model needs to be evaluated with organizations that are concerned with API Management. This is especially important due to the fact that, initially, the Focus Area Maturity Model is populated with practices and capabilities which were extracted from literature. Considering the final purpose of the model, which is for organizations to be able to evaluate, improve upon and assess the degree of maturity their business processes regarding API Management have, these academically grounded practices and capabilities must first be verified through expert interviews with API architects and designers. In doing so, it is likely that additional practices and capabilities that are not mentioned across literature on API Management will be uncovered.

%% file: 6Conclusion.tex
\section{Conclusion} \label{Section:6Conclusion}

API Management is a topic that that has been gaining in popularity in recent years. However, despite growing interest in the subject, more research is needed order to fill knowledge gaps and identify best practices regarding the subject. This is highlighted by the observation that currently, relatively little literature on API Management exists. As a result, no frameworks capturing all the practices, capabilities and features API Management is comprised of exist and an uniform, comprehensive and widely accepted definition of of the topic is lacking within the research community.\\ In order to address these concerns, this document described the execution of a systematic literature review which was conducted in the field of API Management. The purpose of this work is three-fold: first, a comprehensive overview body of literature related to the topic and the posed research questions was collected. This was done by identifying and analyzing 43 books, research papers, theses and whitepapers from a total of 5152, which were extracted from a list of scientific libraries. After having done so, an overview of definitions for API Management was constructed out of the final body of literature, which consists of 24 unique definitions. As a result of a key term frequency analysis using these definitions as input, a new and comprehensive definition of the topic was proposed. Lastly, a collection of practices and capabilities related to API Management was composed. This was done by by scanning and coding the body of included literature. Among the 32 papers that were found to contain at least one practice or capability, 114 practices and 39 capabilities were identified and extracted.

%% file: 7Appendices.tex
\pagebreak
\appendix
\addcontentsline{toc}{section}{Appendices}
\section*{Appendices}

\scriptsize

\section{API Management Definitions}

\setlength\LTleft{0cm}
\setlength{\tabcolsep}{2pt}
\scriptsize
\begin{longtable}{|p{3cm}|p{9.5cm}|p{2cm}|}
\hline
\textbf{Author} & \textbf{Definition} & \textbf{Source}\\\hline
 \hline\hline
Andreo \& Bosch (2019) & In this paper we consider the term API management as described by Jan Stafford\footnote{\url{https:// searchmicroservices.techtarget.com/feature/Why-use-new-lifecycle-tools-inAPI-management-platforms}} where he compares, the API management to application life-cycle management: observing and controlling an API from creation to retirement. &~\cite{andreo2019api}  \\ \hline
Biehl (2015)  &  An API platform typically consists of at least the following three platform components:
API Development Platform: This platform enables API providers to develop APIs quickly and with high quality. It offers API building blocks, which are proven, reusable and configurable. It also offers tools for development and design of APIs. API Runtime Platform: This platform primarily executes the APIs. It serves API responses for incoming API requests of the consumers with favorable non-functional properties, such as high throughput and low latency. API Engagement Platform: This platform allows API providers to manage their interaction with API consumers. It offers API documentation, credentials and rate plans for API consumers. For API providers it offers product management and configuration capabilities. &~\cite{biehl2015api}  \\ \hline
Coste \& Miclea (2019)  &  API management is the practice an organization 
implements to manage the APIs they expose. This is done internally or externally so it makes sure that the APIs are consumable, secure, and available to consumers in conditions agreed in the terms of use of APIs. &~\cite{coste2019api}  \\ \hline
De (2017)  &  An API management platform helps an organization publish APIs to internal, partner, and external developers to unlock the unique potential of their assets. It provides the core capabilities to ensure a successful API program through developer engagement, business insights, analytics, security, and protection. &~\cite{ala2019application,buidesign,de2017api,santana2019case}  \\ \hline
Familiar (2015)  &  API management provides API gateway services such as creating API proxies, configuring SSL and authentication, developer subscription key management, policy injection such as rate throttling or message transformations from JSON to XML or XML to JSON, and performance and health analytics. &~\cite{familiar2015iot}  \\ \hline
Fremantle (2015)  &  One of the key aims of API Management is a desire to manage the following aspects 
orthogonally from the creation of the API itself: – Publishing details of the APIs, documentation, SDKs and other human- and machine-readable material in a portal aimed at developers. 
– Allowing developers to sign up, define application clients, test out Web APIs, and subscribe to them. 
– Managing access control and authentication of API clients using “API keys” or tokens. – Usage control and throttling of traffic to specific clients based on a Service Level Agreement (SLA) or other factors. 
– Monitoring the usage of specific clients in order to be able to limit access or charge for API usage. &~\cite{fremantle2015web}  \\ \hline
Gamez (2015)  & API management features are offered by an API gateway, such as security (e.g. ac- 
cess control, DoS attacks blocking), pricing plans support, API analytics, request monitoring, API lifecycle control (e.g. service orchestration, govern data flows, API design support) or response transformation (e.g. JSON to SOAP). &~\cite{gamez2015towards}  \\ \hline
Hamaleine. (2019)  & API management platforms contain features for monitoring and managing APIs. Usually they include tools for deploying and publishing APIs, and restricting, documenting and versioning them. In addition to these most common management features, API management platforms can have other tools such as management of users, pricing and policies. &~\cite{haselbock2018microservice}  \\ \hline
Haselb{\"o}ck et al. (2018)  & API management is divided into API design, infrastructure aspects, and organizational aspects. API design contains topics such as API specification, API analysis, and API access management. Infrastructure aspects include the tools and frameworks required to deliver APIs to internal and external users. Organizational aspects encompass the organizational tasks required to manage the APIs and responsibilities for managing these tasks &~\cite{haselbock2018microservice}  \\ \hline
Hohenstein et al. (2018)  & Wikipedia \footnote{\url{https://en.wikipedia.org/wiki/API_management}} defines API Management as ``the process of creating and publishing Web APIs, enforcing their usage policies, controlling access, nurturing the subscriber community, collecting and analyzing usage statistics, and reporting on performance.'' &~\cite{haselbock2018microservice}  \\ \hline
Indrasiri \& Siriwardena (2018)  & API management plays a key role in the realization of several microservices governance aspects. The API management layer or API gateway is used to expose microservices to consumers as managed APIs. This includes quality of service aspects as well as several other API management specific details, such as monetization. As part of API management, security, service versioning, throttling, caching and monetization may be applied for services during the runtime. &~\cite{indrasiri2018developing}  \\ \hline
 Kolychev (2019)  & API management systems are comprised 
conventionally of three components: API gateway, API manager, developer site. &~\cite{andrey_kolychev_konstantin_zaytsev_2019_3256462}  \\ \hline
 Liang et al. (2016)  & In terms of API management, we currently see a focus on supplying the API providers side. The increase of available APIs brings the high demand of API management (e.g., API search, API recommendation) in the consumers (developers) side. &~\cite{liang2016exploiting}  \\ \hline
Louren{\c{c}}o Marcos \& Puccinelli de Oliveira (2019) & API management platforms enable API publishing and deployment through a network dashboard. Part of the supported features are policy and access control, back-end guard systems through quotas and rate limits, automatic API deployment, and testing. The dashboard provides additional information on API status, as well as script automation to integrate API deployment into Continuous Integration (CI/CD) pipelines. &~\cite{lourencco2019framework}  \\ \hline
Matsumoto \& Takeda (2017) & API Management is an integrated platform that can manage and release Web-service APIs using abundant functions in a middleware role. It can quickly publish a company’s application APIs to the public and thereby contribute to the digital use of corporate systems. &~\cite{matsumoto2017fujitsu}  \\ \hline
Medjaoui et al. (2018) & API management involves more than just governing the design, implementation, and release of APIs. It also includes the management of an API ecosystem, the distribution of decisions within your organization, and even the process of migrating existing APIs into your growing API landscape. &~\cite{medjaoui2018continuous}  \\ \hline
Nakamura (2017) & API management (provided by a virtual gateway) provides a protocol for each API and function such as billing, load balancing, and activity monitoring, is an effective way of dealing with the complicated calling of diverse APIs. &~\cite{nakamura2017fujitsu}  \\ \hline
Namihira \& Nakajima (2017) & API Management is a service that provides a 
gateway for interconnection by absorbing differences between Web APIs &~\cite{namihira2017iot}  \\ \hline
O'Neil (2014) & The API management solution addresses privacy and security issues through monitoring and visibility capabilities, as well as providing audit trails detailing how its APIs are being used. Essentially, the API management solution will act as a gatekeeper, providing a reliable pathway to control and oversee the flow data between the different connected devices. &~\cite{o2014internet}  \\ \hline
Patni (2017) & An API management tool provides the means to expose an API to external developers in an easy and affordable manner, consisting of the following features: Documentation, Analytics and statistics, Deployment, Developer engagement, Sandbox environment, Traffic management and caching abilities, Security, Availability, Monetization and API lifecycle management. &~\cite{patni2017pro}  \\ \hline
Sine et al. (2015) & API management platforms provide a supplier with a set of services to manage all the constraints 
related to publishing an API. They act somewhat like a portal filtering communications between the API and the applications designed by the external developers. &~\cite{sine2015api}  \\ \hline
Sutherland \& Chetty (2016) & API management software tools automate and control connections between an API and the applications that use it; ensure consistency between multiple API implementations; monitor traffic from individual applications; and protect the API security procedures and policies. &~\cite{sutherland2016investigation}  \\ \hline
Vijayakumar (2018) & API Management is a solution encompassing the collections of tools used to design and manage APIs, referring to both the standards and the tools used to implement API architecture. &~\cite{vijayakumar2018practical}  \\ \hline
Weir (2015) & A basic definition of API Management is the adoption and adaptation of SOA Governance principles and tools in the context of managing the end-to-end lifecycle of an API and the community around it. &~\cite{weir2015oracle}  \\ \hline
 \hline
 \end{longtable}
 \pagebreak

 \section{API Management Capabilities}
\scriptsize
\begin{longtable}{|p{3cm}|p{9.5cm}|p{2cm}|}
\hline
\textbf{Capability} & \textbf{Description} & \textbf{Source}\\\hline
 \hline\hline
Activity Logging  & Activity logging provides basic logging of API access, consumption, performance, and any exceptions. The platform should capture and provide information on who is using an API, what types of apps and devices the API are being called from, and which geographical region is the source of the API traffic~\cite{de2017api}. &~\cite{de2017api,andrey_kolychev_konstantin_zaytsev_2019_3256462}  \\ \hline
Analytics  & The API management platform should be able to extract and log custom variables from within the message payload for advanced analytics reporting. It should provide API administrators and product managers the capability to create pluggable and custom reports from the captured information~\cite{de2017api}. &~\cite{biehl2015api,de2017api,familiar2015iot,indrasiri2018developing,patni2017pro,vijayakumar2018practical,weir2015oracle,hofman2014technical} \\ \hline
API Creation  & The API team should be able to design the REST interface for the API and create an API proxy to interact with the back-end services. An API proxy acts as a facade to securely expose the back-end services to its consumers. Policies attached in the flow paths of the API proxy should be able to implement security, traffic management, message translation, encryption, filtering, caching, orchestration, and routing. Once the development is complete, the API team must be able to deploy and test the API through a console. An embedded console to test APIs can be very handy and can help reduce development time. The API management platform should provide tools that enable the creation of the APIs and subsequently deploy and test them on an environment before they are published for production.~\cite{de2017api} &~\cite{de2017api} \\ \hline
API Discovery & Presence of a mechanism for clients to obtain information about the APIs.~\cite{biehl2015api} &~\cite{biehl2015api,hofman2014technical} \\ \hline
API Proxy & Used in connection with the API mock and functions as communication middle layer between front-end and back-end.~\cite{buidesign} &~\cite{buidesign,familiar2015iot} \\ \hline
API Publication \& Deployment & Once an API has been created, it must be published to an environment before it can be discovered and consumed. The API management platform must therefore provide tools that can be used to migrate the APIs from lower environments and deploy to production.~\cite{de2017api} &~\cite{biehl2015api,de2017api,lourencco2019framework,medjaoui2018continuous,patni2017pro,vijayakumar2018practical,Xu_2019} \\ \hline
Authentication & Authentication is the process of uniquely determining and validating the identity of a client. \cite{de2017api} &~\cite{de2017api,fremantle2015web,gadge2018microservice,gamez2015towards,andrey_kolychev_konstantin_zaytsev_2019_3256462,matsumoto2017fujitsu,montesi2016circuit,vsnuderl2018rate,thielens2013apis,hofman2014technical,Xu_2019} \\ \hline
Authorization & Authorization controls the level of access that 
is provided to an app making an API call. It controls which API resources and methods that an app can invoke.~\cite{de2017api} &~\cite{de2017api,gadge2018microservice,andrey_kolychev_konstantin_zaytsev_2019_3256462,thielens2013apis} \\ \hline
Availability & Ensuring the API is accessible during certain hours.~\cite{preibisch2018api} &~\cite{patni2017pro,preibisch2018api} \\ \hline
Catalog \& Documentation  & Developer enablement services should allow an API provider to publish a discoverable catalog of APIs. An API catalog is also sometimes referred to as an API registry. Developers should be able to search the catalog based on various metadata and tags. The catalog should document the API functionality, its interface, how-to guides, terms of use, reference documents, and so forth. Information about the API versions available should also be included in the documentation.\cite{de2017api} &~\cite{biehl2015api,de2017api,gamez2015towards,jacobson2011apis,lourencco2019framework,patni2017pro,vijayakumar2018practical,hofman2014technical} \\ \hline
Community Management & App developers often like to know the views of other developers in the community. They may want to collaborate and share their API usage learnings and experiences with one another. Blogs and forums form a major part of collaboration and community management.~\cite{de2017api} &~\cite{biehl2015api,de2017api,weir2015oracle} \\ \hline
Consumer Onboarding & \textit{Consumer Onboarding is the process of teaching a new consumer of an API about the way the API works, what it can do and what problems it aims to solve. Ideally, this process should be made to be as easy and streamlined as possible.} &~\cite{biehl2015api} \\ \hline
Developer Onboarding & To start consuming the APIs, developers must register with the API provider to get access credentials. Developers can either sign up independently or as part of a company. The signup process should be simple and easy. Developers should be able to go through a self-registration process and view the APIs available from the API provider.~\cite{de2017api} &~\cite{de2017api,fremantle2015web,medjaoui2018continuous} \\ \hline
Developer Portal Support & A developer portal is a customized web site that allows an API provider to provide services to the developer community. It is essentially a content-management system that documents the APIs—their functionalities, interfaces, getting-started guides, terms of use, and much more. Developers can sign up through the portal and register their applications to use the APIs.~\cite{de2017api} &~\cite{de2017api,familiar2015iot,fremantle2015web,sine2015api} \\ \hline
Developer Support & Properly designed REST APIs are normally very intuitive for developers to understand. App developers can easily start using them for app development. Still, the API provider should provide resources that developers can use to build innovative apps. Good API documentation and accelerators in the form of test and development kits can help speed up the adoption of APIs.~\cite{de2017api} &~\cite{de2017api,jacobson2011apis,medjaoui2018continuous,raivio2011towards} \\ \hline
Encryption & Often, message payloads sent in API calls contain sensitive information that can be the target for man-in-the-middle attacks. An API management platform sits in between the client app and the API service provider as an API gateway. All communication between the client app and the API service provider through the intermediate API gateway should be secured using SSL/TLS encryption by default.~\cite{de2017api}  &~\cite{de2017api,mussig2017highly,preibisch2018api}  \\ \hline
Governance & It is essential to establish policies and monitoring. The policies can broadly be categorized as design-time governance and runtime governance. The policies are highly influenced by IT (business) objectives and goals.~\cite{gadge2018microservice} &~\cite{de2017api,gadge2018microservice,medjaoui2018continuous} \\ \hline
Identity Mediation & APIs normally use OAuth protocols for implementing security. However, the back-end services may be secured using SAML or any other WS-Security headers. Hence, the API management platform must have the capability to integrate with back-end IDM platforms and do identity mediation.~\cite{de2017api} &~\cite{de2017api} \\ \hline
Interface Translation & When an enterprise creates an API to expose its data and services, it needs to ensure that the API interface is intuitive enough for developers to easily use. APIs should be created with an API-First approach, which promotes API creation with a consumer focus. Hence, the interface for the API will most likely be different from that of the back-end services that it exposes. The API gateway should therefore be able to transform the API interface to a form that the back end can understand.~\cite{de2017api} &~\cite{de2017api} \\ \hline
Issue Management & The API management platform should provide API consumers with the facility to log issues found in the APIs. App developers consuming APIs must be able to report any issues or shortcomings related to their APIs. They should be able to raise support tickets and seek help regarding API usage.~\cite{de2017api} &~\cite{de2017api,jacobson2011apis} \\ \hline
Key \& Certificate Management & The API management platform should provide the capability to manage keys and certificates required for data privacy.~\cite{de2017api} &~\cite{de2017api,hohenstein2018architectural,sine2015api,thielens2013apis,gadge2018microservice}  \\ \hline
Lifecycle Management & The API Lifecycle Manager allows you to develop, document, scale, 
and version APIs. It also provides API Lifecycle Management-related tasks, such as publishing an API, monetization, analyzing statistics, and promoting.~\cite{indrasiri2018developing} &~\cite{biehl2015api,indrasiri2018developing,patni2017pro,weir2015oracle,hofman2014technical}  \\ \hline
Monetization & Monetizing an API by charging for additional traffic.~\cite{jacobson2011apis} &~\cite{biehl2015api,de2017api,indrasiri2018developing,jacobson2011apis,patni2017pro,weir2015oracle}  \\ \hline
Monitoring & An API management platform collects a wide variety of operational and business data as traffic flows through it. The data collected is then analyzed to provide metering and monitoring capabilities.~\cite{de2017api} &~\cite{biehl2015api,de2017api,gadge2018microservice,jacobson2011apis,medjaoui2018continuous,montesi2016circuit,thielens2013apis,hofman2014technical,Zhao_2018}  \\ \hline
Quality of Service & API Quality of Service (QoS) has multiple perspectives such as security, caching, throttling, and other SLAs (Service Level Agreements).~\cite{indrasiri2018developing}  &~\cite{indrasiri2018developing,hofman2014technical}  \\ \hline
Scaling Management & \textit{Scaling management is concerned with (automatically) scaling the amount of available resources up or down depending on an APIs usage.}  &~\cite{akbulut2019software,jacobson2011apis,gadge2018microservice,hofman2014technical}  \\ \hline
Security & APIs provide access to valuable and protected data and assets. Therefore, security for APIs is of utmost importance to protect the underlying assets from unauthenticated and unauthorized access.~\cite{de2017api}  &~\cite{de2017api,gadge2018microservice,jayathilaka2015eager,montesi2016circuit,medjaoui2018continuous,patni2017pro,vijayakumar2018practical,hofman2014technical,Xu_2019}  \\ \hline
Service Discovery & The number and location of service instances is dynamic. Consequently, client code needs to use a more elaborate service discovery mechanism. There are two main service discovery patterns: client-side discovery and server-side discovery.~\cite{gadge2018microservice}  &~\cite{gadge2018microservice,indrasiri2018developing,montesi2016circuit,Xu_2019}  \\ \hline
Service Orchestration & In many scenarios, the API gateway may need to invoke multiple back-end services in a particular sequence or in parallel and then send an aggregated response to the client. This is known as service orchestration. The service orchestration capability helps to create a coarse-grained service by combining the results of multiple back-end services invocation.~\cite{de2017api}  &~\cite{de2017api,gadge2018microservice,gamez2015towards,preibisch2018api}  \\ \hline
 Service Registry & The service registry helps to keep track of these instances. It is a database containing the network locations of the service instances. Every service instance registers itself on start-up and de-registers on shutdown. The API Gateway uses this information during service discovery.~\cite{gadge2018microservice}  &~\cite{gadge2018microservice,indrasiri2018developing}  \\ \hline
 Service Routing & APIs need to route requests from consumers to the right back-end service providing the business functionality. There may be one more backend systems providing the backend functionality. Hence,the API management platform should be able to identify and route the request to the correct instance of the back-end.~\cite{de2017api}  &~\cite{akbulut2019software,ciavotta2017microservice,de2017api,mussig2017highly,preibisch2018api}  \\ \hline
Service-Level Agreements (SLA) & A service-level agreement (SLA) defines the API’s non-functional requirements. This can include the expected throughput, response time, rate limits for various tiers (if applicable), maintenance or downtime information, and so forth.~\cite{de2017api}  &~\cite{biehl2015api,de2017api,jacobson2011apis,vijayakumar2018practical}  \\ \hline
Service-Level Monitoring & The API management platform should provide performance statistics that track the latency within the platform and the latency for back-end calls. This helps the API administrator find the source of any performance issues reported on any API.\cite{de2017api}  &~\cite{de2017api,Xu_2019}  \\ \hline
Subscription Management & \textit{Subscription Management is the activity where the publisher of an API manages existing subscriptions (consumers of) on the API.}   &~\cite{fremantle2015web,raivio2011towards}  \\ \hline
Testing & Often, APIs are exposed via an API management platform configured to implement 
different policies on top of the API business logic. These policies may implement security, traffic throttling, data transformation, routing, or orchestration. API testing strategy must consider testing these policies in the API gateway.\cite{de2017api}  &~\cite{de2017api,familiar2015iot,jacobson2011apis,patni2017pro,medjaoui2018continuous}   \\ \hline
Threat Protection & To protect its APIs from different types of threats, an organization must build an API proxy in front of the APIs with an API management platform and implement security policies in these proxies to protect against such threats.~\cite{de2017api}  &~\cite{de2017api,preibisch2018api,sine2015api}   \\ \hline
Traffic Management & Depending on the nature of data and services provided by the API, traffic management offers a different business value to different classes of customers. Each customer class may be willing to pay differently for access.~\cite{de2017api}  &~\cite{de2017api,preibisch2018api,sine2015api}  \\ \hline
Transformation & The API Gateway provides a place for data transformation, where messages can be translated between backend, API, and app formats and protocols. The gateway provides a central data transformation point through which all traffic is translated for: 
- Requested payload transformations - Header transformations - Protocol transformations~\cite{gadge2018microservice}  &~\cite{gadge2018microservice,andrey_kolychev_konstantin_zaytsev_2019_3256462,preibisch2018api}  \\ \hline
Version Management & APIs evolve over time with newer business requirements. Hence, managing multiple versions of an API to support existing consumers is an important capability that must be provided by the API management platform. Version management should also provide the ability to deprecate and retire older versions smoothly.~\cite{de2017api}  &~\cite{akbulut2019software,biehl2015api,de2017api,medjaoui2018continuous,gamez2015towards,vijayakumar2018practical}  \\ \hline
 \hline
 \end{longtable}
 
 \pagebreak

 \section{API Management Practices}
\scriptsize
\begin{longtable}{|p{3cm}|p{9.5cm}|p{2cm}|}
\hline
\textbf{Practice} & \textbf{Description} & \textbf{Source}\\\hline\hline
Access Control & Testing access mechanism and access control policies of an API is of paramount importance. If an API is exposing a protected resource, the security testing must ensure that only authenticated clients are able to access the APIs.~\cite{de2017api}  &~\cite{de2017api,gadge2018microservice,nakamura2017fujitsu,raivio2011towards,sine2015api,vijayakumar2018practical}  \\ \hline
Access Token Authorization & An access token is generated as part of the OAuth handshake and is associated with scopes that determine the APIs that can be accessed using the token. An access token can be associated with one or multiple scopes. Each access token may have an expiry duration that controls the duration for which the token is valid.~\cite{de2017api}  &~\cite{de2017api,mussig2017highly,patni2017pro,Zhao_2018}  \\ \hline
Activity Logging/Monitoring & Activity logging provides basic logging of API access, consumption, performance, and any exceptions.~\cite{de2017api}  &~\cite{biehl2015api,de2017api,jacobson2011apis,fremantle2015web,lourencco2019framework,preibisch2018api,sine2015api,thielens2013apis}  \\ \hline
Adopt API Symbol Labeling Method & In cases where API calls represent unavoidable risk, you can reduce negative impacts by adding warnings to the API documentation itself. A way to warn API users is to adopt a labeling method using symbols. This way, there is no need to add lots of text to your documentation: readers can just recognize the warning label instead.~\cite{medjaoui2018continuous}  &~\cite{medjaoui2018continuous}  \\ \hline
AppKey Authentication & Every app is identified by its name and a unique UUID known as the app key. The app key often serves as an identity for the app making a call to the API. It is normally issued and managed via the API management platform of the API provider. An app key is also known as an API key, an app ID, or a client ID. The API management platform must have the ability to issue, track, and revoke the app key.~\cite{de2017api}  &~\cite{biehl2015api,de2017api,Zhao_2018}  \\ \hline
Authenticated User Rate Limitation & The maximum number of requests an API can access within a given time frame may be based on the rate limiting approach. This can be accomplished by deciding that an API may be accessed only once a day, by an authenticated user from a specific application.~\cite{gadge2018microservice}  &~\cite{gadge2018microservice}  \\ \hline
Automatic API Deployment & Automatic API deployment may be achieved by using script automation to integrate API deployment into Continuous Integration (CI/CD) pipelines.~\cite{lourencco2019framework}  &~\cite{lourencco2019framework}  \\ \hline
Backwards Compatibility Checking & In addition to the governance policy validations, EAGER 
also performs a set of built-in sanity checks on all applications and APIs deployed in the PaaS cloud. One of these checks is the backwards compatibility check. If EAGER detects that an API which is already deployed in the cloud is being redeployed, it performs a comparison between the old and latest specifications of the API to make sure that the developer is not introducing a backward incompatible change to the API.~\cite{jayathilaka2015eager}  &~\cite{jayathilaka2015eager}  \\ \hline
Billing & Billing is another important aspect of API monetization. Once APIs have been monetized, it is the necessary to generate consumer bills at regular intervals. Some API management platforms provide an integrated billing solution that automatically generates billing documents such as invoices and revenue share statements at prescheduled intervals.~\cite{de2017api}  &~\cite{de2017api,nakamura2017fujitsu,raivio2011towards,sine2015api,hofman2014technical}   \\ \hline
Black or Whitelist IP Adresses & Whitelist policies such as using IP address and address range throttles access for a given IP address or address range. This is useful to give internal users (with internal IPs) other quotas than for external ones. Blacklists are also possible to exclude specific IPs from any request.~\cite{hohenstein2018architectural}  &~\cite{gadge2018microservice,gamez2015towards,hohenstein2018architectural}   \\ \hline
Blocking Requests & Most API Gateways provide features that can handle DDoS attacks, by regulating and controlling the traffic as it proceeds to backend microservices. Among other traffic regulating parameters, the API Gateway may be configured to block requests that: - seem to target a specific API that have a User-Agent header, set to a value that does not correspond to normal client traffic. - Have a referrer header, set to a value that can be associated with an attack. 
- Have other headers with values that can be associated with an attack.~\cite{gadge2018microservice}  &~\cite{gadge2018microservice}   \\ \hline
Business Value Reporting & Business value reports gauge the monetary value associated with the API program. Monetization reports of API usage provide information on the revenue generated from the API. The API gateway should be able to provide API usage monetization reports.~\cite{de2017api}  &~\cite{buidesign,de2017api}   \\ \hline
Caching & Caching is a mechanism to optimize performance by responding to requests with static responses stored in-memory. An API proxy can store back-end responses that do not change frequently in memory. As apps make requests on the same URI, the cached response can be used to respond instead of forwarding those requests to the back-end server. Thus caching can help to improve an API’s performance through reduced latency and network traffic.~\cite{de2017api}  &~\cite{biehl2015api,de2017api,gadge2018microservice,gamez2015towards,indrasiri2018developing,patni2017pro,preibisch2018api,vsnuderl2018rate,vijayakumar2018practical,hofman2014technical,Zhao_2018}    \\ \hline
 Change Notifications & Changes to an API may adversely affect its consumers. Hence, consumers must be notified of any planned changes to the API. Developers using the APIs should be made aware of any changes to the API. The API management platform must therefore provide a mechanism to notify API consumers of any API upgrades or outages. Notification can be made via email, SMS, or social media.~\cite{de2017api}  &~\cite{de2017api}    \\ \hline
Circuit Breaking & An API gateway may offer an implementation of the Circuit Breaker pattern, impeding the system to get stuck in case the target back-end service fails to answer within a certain time.~\cite{ciavotta2017microservice}  &~\cite{akbulut2019software,ciavotta2017microservice,montesi2016circuit}    \\ \hline
Client SDK Generation & \textit{Software Development Kits (SDKs) contain the toolkits which are necessary to create a client application to invoke a particular API. If an API consumer wants to create an application,an API Management platform may offer them the option to generate a client side SDK for a supported language or framework. They may then use this SDK to write a software application which consumes the APIs of their choosing.} &~\cite{vijayakumar2018practical}    \\ \hline
Closing Slow Connections & Most API Gateways provide features that can handle DDoS attacks, by regulating and controlling the traffic as it proceeds to backend microservices. Among other traffic regulating parameters, the API Gateway may be configured to close slow connections. In doing so, after a certain time span a connection will be closed from a client that is writing data too infrequently, which can represent an attempt to keep connections open as long as possible.~\cite{gadge2018microservice}  &~\cite{gadge2018microservice}    \\ \hline
Cloud-Based API Gateway & As more API services are sourced and distributed in the cloud, the concept of CDNs can also apply to APIs. API gateways can be placed in the cloud and act as an intermediate proxy layer, caching some traffic before it ever gets to your network.~\cite{jacobson2011apis}  &~\cite{jacobson2011apis}    \\ \hline
Clustering & \textit{Clustering allows developers to configure multiple nodes that connect to a single database. Using load balancing, this cluster is able to present a unified network interface to client systems and software. All nodes in the cluster will share the same configuration, due to the fact that they point to the same database.}  &~\cite{hofman2014technical}    \\ \hline
Community Blog Support & API providers may provide aids for developers in building and monitoring their apps. A developer community may consist of blogs to support developers and help them collaborate.~\cite{de2017api}  &~\cite{biehl2015api,de2017api,raivio2011towards}    \\ \hline
Community Forum Support & API providers may provide aids for developers in building and monitoring their apps. A developer community may consist of a forum to support developers and help them collaborate.~\cite{de2017api}  &~\cite{biehl2015api,de2017api,raivio2011towards}    \\ \hline
Connection Pooling & The API management platform should be able to maintain a pool of connections to the back-end service. Connection pooling improves overall performance. Also, it may be required for traffic management purposes to ensure that only a fixed maximum number of active connections are opened at any point in time to the back-end service.~\cite{de2017api}  &~\cite{biehl2015api,de2017api}    \\ \hline
Create API Endpoints & \textit{The place that APIs send requests to and where the resource they need to carry out their function are located, is called an endpoint. An API Management platform may offer users the ability to create endpoints for their APIs.} &~\cite{vijayakumar2018practical}    \\ \hline
Creating API Proxies & The API team should be able to design the REST interface for the API and create an API proxy to interact with the back-end services. An API proxy acts as a facade to securely expose the back-end services to its consumers.~\cite{de2017api}  &~\cite{de2017api,familiar2015iot}    \\ \hline
Data Masking & APIs expose data that may be sensitive; such data should be visible only to its intended recipient. If such data gets logged anywhere, it must be masked. Masking sensitive data at rest within audits and log files is yet another data privacy requirement that an API management platform should provide.~\cite{de2017api}  &~\cite{de2017api}    \\ \hline
Data Privacy & The API management platform must possess data privacy capabilities. Data privacy can be achieved through encryption and data masking.~\cite{de2017api}  &~\cite{de2017api}    \\ \hline
Data Transformation & \textit{An API gateway may provide users with a mechanism to add data transform rules to their API. Data transformations may include operations such as URL rewriting or transforming XML to JSON. } &~\cite{de2017api,vijayakumar2018practical}    \\ \hline
 Decouple API \& Software Versioning & Every major release, enhancements, and bug fixes result in a new version of the software. If we start tying the API version to its software implementation, it would result in an unprecedented number of API versions. This would not only frustrate the consumers dealing with the API, but also results in maintenance nightmares for API providers. Hence, the API version should never be tied to the software version of the back-end data/service. A new API version should be created only if there is a change in the contract of the API that impacts the consumer.~\cite{de2017api} &~\cite{de2017api} \\ \hline
Define Migration Period & A date for deprecation of the older version of the API should be set so that developers have a clear target to migrate their apps over to the new API version.~\cite{de2017api}  &~\cite{de2017api}    \\ \hline
Design Time Governance & A Design time governance consists of: - Defining the standards for API definitions (example: Swagger) - Keyword tags used to categorize APIs  - Conformance to REST API design guidelines~\cite{gadge2018microservice}  &~\cite{gadge2018microservice}    \\ \hline
Developer Key & A developer key must be included in the API call in order for the call to pass 
through to the actual API. That means that anyone who wants to invoke your API must first register on the Developer Portal and request to subscribe to the API. Once they are approved, they receive a key. The developer key is the element that allows the service to gather statistics on who is calling, how often, and what performance those invocations are receiving.~\cite{familiar2015iot}  &~\cite{familiar2015iot}    \\ \hline
Developer Portal Announcements & New upcoming version and versioning schedules should be announced in the API developer portal.~\cite{de2017api}  &~\cite{de2017api}    \\ \hline
Display Deprecation Warning & Introduce “warning” headers in alerts on older versions being deprecated. \cite{de2017api}  & \cite{de2017api}    \\ \hline
DoS Protection & Hackers may try to bring down back-end systems by pumping unexpectedly high traffic through the APIs. Denial-of-service (DoS) attacks are very common on APIs. Hence, the API management platform should be able to detect and stop such attacks. \cite{de2017api}  &~\cite{de2017api,gadge2018microservice,gamez2015towards}    \\ \hline
Email Versioning Notification & Emails should be sent to registered developers about upcoming new API versions.~\cite{de2017api}  &~\cite{de2017api}    \\ \hline
Email-Based Support & The API support information in the portal should provide the developers’ contact information so as to reach in case of any queries or issues with using the API. The contact information can be a phone number or an email address.~\cite{de2017api}  &~\cite{de2017api,jacobson2011apis}    \\ \hline
Error Handling & \textit{Error handling refers to the way in which an API Management platform handles and presents errors or exceptions to the user when one occurs during consumption of an API.}  &~\cite{de2017api,andrey_kolychev_konstantin_zaytsev_2019_3256462}    \\ \hline
Error Reporting & APIs should be monitored closely so that any errors that occur can be recorded and reported.~\cite{medjaoui2018continuous}  &~\cite{medjaoui2018continuous}    \\ \hline
Escalation Plan & \textit{An escalation plan refers to the strategy an API provider has in place with regards to the event where the consumption quota of the API is exceeded.} &~\cite{jacobson2011apis}    \\ \hline
Federated Identity Support & Federated identity is the preferred solution for implementing authentication and authorization in microservice architecture. Each microservice does not necessarily need to obtain and store users’ credentials in order to authenticate them. Instead, microservices can use an identity management system that is already storing a user’s identity to authenticate the user. This approach allows the decoupling of the authentication and authorization functions.~\cite{gadge2018microservice} &~\cite{gadge2018microservice}    \\ \hline
Filtering Spam Calls & \textit{Functionality offered by the API Management platform with which duplicate/spam/excessive amounts of calls to an API are filtered out.} &~\cite{akbulut2019software}    \\ \hline
Format Translation & The back-end system might expect data in SOAP, or XML, or CSV or any other proprietary format. Such data format cannot be easily consumed by the API consumer. Hence, the API gateway should have the capability to easily transform from one format to other.~\cite{de2017api} &~\cite{de2017api}    \\ \hline
Form-Based Support & Support through forms filled in on a webpage.~\cite{jacobson2011apis} &~\cite{jacobson2011apis}    \\ \hline
Fuzzy-Based Scaling & \textit{Scaling technique which utilizes fuzzy logic.} &~\cite{akbulut2019software}    \\ \hline
Health Monitoring & Health monitoring is done to make sure the gateway is up and running. For health monitoring, it is recommended to capture: - System status and health (CPU, Memory, Thread usage) - Network connectivity - Security alerts - Backups and recovery - Maintenance of logs~\cite{gadge2018microservice} &~\cite{gadge2018microservice}    \\ \hline
Host Name Versioning & Another approach to API versioning using a URL is to use a different host name. For example, Facebook’s first version of an API is available at api.facebook.com, whereas their new graph API is available at graph.facebook.com. This approach is used only when there is an extensive revamp of the API.~\cite{de2017api} &~\cite{de2017api}    \\ \hline
HTTP Header Versioning & Another approach to API versioning is to use an HTTP header. With this approach, an HTTP header is used by the client to specify the API version that it wants to invoke.~\cite{de2017api} &~\cite{akbulut2019software,de2017api}    \\ \hline
Interactive \& Automatically Generated Documentation & \textit{Functionality offered by an API Management platform with which API providers are able to automatically generate documentation for their API.} &~\cite{biehl2015api}    \\ \hline
Interactive API Console & \textit{Component of an API Management platform (usually as part of the developer portal) with which developers are able to test the behavior of an API.} &~\cite{biehl2015api}    \\ \hline
JSON Threat Protection Policies & JSON threat protection policies can be used to check the message payload for the 
following, and reject the message if any of the allowed limits are exceeded.~\cite{de2017api} &~\cite{de2017api}    \\ \hline
JWT Authentication & An API gateway may use an identity management and authentication service which manages accounts, such as JWT, to allow a user or client access to certain microservices.~\cite{Xu_2019} &~\cite{Xu_2019}    \\ \hline
Load Balancing & Load balancing helps to distribute API traffic to the back-end services. Various load balancing algorithms may be supported. Based on the selected algorithm, the requests must be routed to the appropriate resource that is hosting the service. Load balancing capabilities also improve the overall performance of an API.~\cite{de2017api} &~\cite{biehl2015api,ciavotta2017microservice,de2017api,gadge2018microservice,gamez2015towards,montesi2016circuit,nakamura2017fujitsu,Xu_2019,Zhao_2018}    \\ \hline
Maintain Multiple API Versions & The API provider has to maintain multiple versions of the API for a sufficient time in order to enable a smooth transition. Supporting multiple versions requires significant investment by both the API provider and the consumers.~\cite{de2017api} &~\cite{de2017api}    \\ \hline
Manual Scaling & \textit{Scaling technique with which users of an API Management platform may manually scale the amount of available resources up or down depending on API usage.} &~\cite{akbulut2019software}    \\ \hline
Minimize API Updates Frequency & Every time a new API version is released, it kick-starts a fresh cycle for app developers. They need to understand the new API, analyze the impact on their apps, debug issues, and so on. This is a huge burden on both sides in terms of time and money. Because of this, it is advised to keep the frequency of major API versions to a minimum.~\cite{de2017api} &~\cite{de2017api}    \\ \hline
OAuth + JWT Authentication & The OAuth2 + JWT process is exactly the same as OAuth2. OAuth2 will eventually issue an Access Token to the caller. OAuth2+JWT actually replaces the Access Token with the JWT. The benefit of doing so is simply to reduce the number of queries to the DB at the time of the Token check.~\cite{Zhao_2018} &~\cite{Zhao_2018}    \\ \hline
OAuth Authentication & OAuth 2.0 is a protocol that allows clients to grant access to server resources without sharing credentials. As per the IETF specifications, the OAuth 2.0 authorization framework enables a third-party application to obtain limited access to an HTTP service, either on behalf of a resource owner by orchestrating an approval interaction between the resource owner and the HTTP service, or by allowing the third-party application to obtain access on its own behalf.~\cite{de2017api} &~\cite{de2017api,gadge2018microservice,gamez2015towards,hohenstein2018architectural,matsumoto2017fujitsu,patni2017pro,thielens2013apis,hofman2014technical,Xu_2019,Zhao_2018}    \\ \hline
OAuth to SAML Identity Mediation & APIs normally use OAuth protocols for implementing security. However, the back-end services may be secured using SAML or any other WS-Security headers. Hence, the API management platform must have the capability to integrate with back-end IDM platforms and do identity mediation. OAuth to SAML is a very common identity mediation requirement.~\cite{de2017api} &~\cite{de2017api}    \\ \hline
Offloading & \textit{Offloading is a gateway pattern which is concerned with moving shared or specialized service functionality to a gateway proxy. This pattern can simplify application development by moving shared service functionality, such as the use of SSL certificates, from other parts of the application into the gateway.} &~\cite{akbulut2019software}    \\ \hline
Open Source Support & \textit{Type of gateway or management platform which provides its users with the option to extend and customize functionalities due to its open source nature.} &~\cite{gadge2018microservice,hofman2014technical}    \\ \hline
OpenID Connect Authentication & OpenID Connect 1.0 is an authentication protocol that builds on top of OAuth 2.0 specs to add an identity layer. It extends the authorization framework provided by OAuth 2.0 to implement authentication.~\cite{de2017api} &~\cite{de2017api,gadge2018microservice}    \\ \hline
Operations Runbook & The API team also needs documentation. Operations, in particular, needs a “runbook” that details responses to common problems. This document can help you move from an “all hands on deck” start-up to a smoothly run, professionally managed operation.~\cite{jacobson2011apis} &~\cite{bui,jacobson2011apis}    \\ \hline
Phone-Based Support & \textit{Type of support where API providers provide support to their consumers, using phone calls as a communication medium.}  &~\cite{jacobson2011apis}    \\ \hline
Policy Control & \textit{Policies allow API providers to change the behavior of an API through configuration. An API Management platform may provide providers with the ability to control these policies.} &~\cite{familiar2015iot,lourencco2019framework}    \\ \hline
Protocol Translation & API management platforms must be able to do a protocol transformation from SOAP to REST to provide a lightweight interface for consumers. Support for other protocol transformations—like HTTP(s) to JMS/FTP/JDBC—may be a nice to have feature in the API management platform.~\cite{de2017api} &~\cite{de2017api,gamez2015towards,matsumoto2017fujitsu,vijayakumar2018practical}    \\ \hline
Provide API Status Page & An API status page is a special webpage that is supported on different technical infrastructure than the main API and whose only function is to let users know what’s going on with the API at a technical level at any time they choose to have a look.~\cite{jacobson2011apis} &~\cite{jacobson2011apis}    \\ \hline
Provide Source Code Example & \textit{An API Management platform may provide developers with source code examples that are meant to be re-used, provide inspiration or serve as learning aids.} &~\cite{biehl2015api}    \\ \hline
Query Parameter Versioning & This is yet another common approach to versioning API. In this approach the client specifies the version number as a query parameter in the request, as follows: http://www.foo.com/customers?version=v2 The server may choose to honor the query parameter or even ignore it. One 
advantage of this option is that the version parameter can be optional or required, depending on how you want the API to be used.~\cite{jacobson2011apis} &~\cite{jacobson2011apis}    \\ \hline
Rate/Quota Limiting & With a Rate Limiting approach (also sometime referred to as Quota), the requests are throttled based on the originating app or user, region of origination, time of the day and various other factors over a period of time. The request within the specified limit is routed successfully to the target system. Those beyond the limit are rejected.~\cite{de2017api}  &~\cite{de2017api,gamez2015towards,jacobson2011apis,lourencco2019framework,raivio2011towards,jayathilaka2015eager,vsnuderl2018rate,hofman2014technical,gadge2018microservice}    \\ \hline
Reference Documentation & Provide reference documentation that may be auto-generated for completeness; it should document every API call,every parameter, and every result so that developers can be 100 percent clear on how the API functions.~\cite{jacobson2011apis}  &~\cite{jacobson2011apis,medjaoui2018continuous}    \\ \hline
Refresh Token Authorization & Refresh tokens represent a limited right to reauthorize the granted access by obtaining new access tokens.~\cite{de2017api}  &~\cite{de2017api,patni2017pro}    \\ \hline
Request Origin Rate Limitation & The maximum number of requests an API can access within a given time frame may be based on the rate limiting approach. To do so, the request origin approach my be used.~\cite{gadge2018microservice}  &~\cite{gadge2018microservice}  \\ \hline
Role-Based Access Control & Employees are permitted to conduct certain transactions based on their roles.~\cite{thielens2013apis}  &~\cite{de2017api,thielens2013apis,hofman2014technical}  \\ \hline
Rule-Based Scaling & \textit{Scaling technique which is based on a set of rules and conditions.} &~\cite{akbulut2019software}    \\ \hline
Runtime Policy Governance & Runtime governance consists of tracking the life cycle of APIs: - Handling routing, blocking, and processing  - Understanding the API utilization and raising the alerts/ alarms in case usage crosses the threshold  - Traffic throttling, smoothing, and load balancing  - Rate limiting per-API usage - API versioning - Schema versioning for input/ output request parameters.~\cite{gadge2018microservice} &~\cite{gadge2018microservice,jayathilaka2015eager}    \\ \hline
SAML Authentication & \textit{SAML, the Security Assertion Markup Language, is an XML-based identity federation standard.} &~\cite{de2017api,gadge2018microservice,hofman2014technical}    \\ \hline
Sandbox Environment Support & Adds dummy values to mock an API based on a provided specification. ~\cite{buidesign} &~\cite{buidesign,jacobson2011apis,patni2017pro}    \\ \hline
Service \& Data Mapping & An API management platform should provide a graphical representation of the different back-end service component that maps to provide an API service. It should incorporate service mapping tools that enable the discovery and description of existing service delivery assets so that they can be wired into your API design. ~\cite{de2017api} &~\cite{de2017api}    \\ \hline
Service Dispatching & This allows the API management platform to select and invoke the right back-end service. In some cases, multiple services may have to be invoked to perform some sort of orchestration and return an aggregated response to the consumer. ~\cite{de2017api} &~\cite{de2017api}    \\ \hline
Spike Arrest & Identifies an unexpected rise in the API traffic. It helps to protect back-end systems that are not designed to handle a high load. API traffic volume exceeding the spike arrest limit may be dropped by the API management platform to protect back-end systems in the event of DoS attacks. ~\cite{de2017api} &~\cite{de2017api,jacobson2011apis}    \\ \hline
SQL Injection Protection & It is important that any API that accepts input that can be inserted into an 
SQL database must be protected against SQL injection attacks. Regular expressions that match certain SQL keywords can be used to detect malicious SQL content in the API request. ~\cite{de2017api} &~\cite{de2017api,preibisch2018api}    \\ \hline
SSL Encryption & Sensitive data should be encrypted with digital certificates in transit. The API management platform should have support for SSL/TLS. For some use cases, additional encryption of specific elements within the payload may also be required.~\cite{de2017api} &~\cite{de2017api,familiar2015iot,hofman2014technical}    \\ \hline
Startup Documentation & This type of documentation explains key concepts and accelerates understanding. Without this, developers may flail through the reference documentation without being able to make connections that could appear obvious to insiders.~\cite{jacobson2011apis} &~\cite{jacobson2011apis}    \\ \hline
Support Request Prioritization & Prioritize support requests so that the biggest issues and most important customers are given priority.~\cite{jacobson2011apis} &~\cite{jacobson2011apis}    \\ \hline
Support Request Tracking & Track support requests to ensure that they are not forgotten or ignored~\cite{jacobson2011apis} &~\cite{jacobson2011apis}    \\ \hline
Threat Detection & The API management platform should be able to identify malformed request formats or malicious content within the payload and then protect against such attacks. Error visualization capability can also help detect any hacker attempting to find an exploitable weakness in APIs.\cite{de2017api} &~\cite{de2017api}    \\ \hline
Tier-Based Monetization & The desire to sell premium access to the API, creating service tiers.~\cite{jacobson2011apis} &~\cite{jacobson2011apis,hofman2014technical}    \\ \hline
TLS Encryption & Sensitive data should be encrypted with digital certificates in transit. The API management platform should have support for SSL/TLS. For some use cases, additional encryption of specific elements within the payload may also be required. \cite{de2017api} &~\cite{de2017api,preibisch2018api}    \\ \hline
Token Translation \& Management & \textit{(Access) token management is an activity that involves measures to securely obtain, store, and manage the security tokens which are required to invoke backend APIs.} &~\cite{hofman2014technical}    \\ \hline
Traffic Prioritization & Helps the API management platform determine which class of customers should be given higher priority. API calls from high-priority customers should be processed first. Not all API management platforms support this capability.~\cite{de2017api} &~\cite{de2017api}    \\ \hline
URI Versioning & This approach is semantically meaningful since it uses the version information in the Uniform Resource Identifier (URI). A simple example of this might look like http://v1.example.com/service/ or http://api.example.com/service/v1/. The representation of an API is immutable, and a fresh URI space needs to be created, such as, http://api.example.com/service/v2/, with the publication of a new version.~\cite{akbulut2019software} &~\cite{akbulut2019software}    \\ \hline
URL Mapping & The path of the incoming URL may be different from that of the back-end service. A URL mapping capability allows the platform to change the path in the incoming URL to that of the back-end service. This URL mapping happens at runtime so that the requested resource is retrieved by the consumer via service dispatching.~\cite{de2017api} &~\cite{de2017api}    \\ \hline
URL Rewriting & Before the reverse proxy is performed, an instance is obtained by polling based on the instance information of the API. In addition, according to the rewriting rules of the URL, the original URL is converted into the URL of the actual internal service, and the call address of the reverse proxy is generated in conjunction with polling the selected instance address.~\cite{Zhao_2018} &~\cite{vijayakumar2018practical,Zhao_2018}    \\ \hline
URL Versioning & An API is normally identified by its URL. So in API versioning, it makes sense to introduce the version information in the URL as follows: 
http://www.foo.com/v1/customers In this URL, v1 defines the major version identifier. When this identifier changes, it is assumed that all resources under it changes—in this case, the customers resource. If a new version of the customer is introduced in the next version, it should use a new version identifier, like /v2/customers.~\cite{de2017api} &~\cite{de2017api}    \\ \hline
Usage Throttling & Provides a mechanism to slow down subsequent API calls. This can help to improve the overall performance and reduce impacts during peak hours. It helps to ensure that the API infrastructure is not slowed down by high volumes of requests from a certain group of customers or apps.~\cite{de2017api} &~\cite{de2017api,fremantle2015web,familiar2015iot,gadge2018microservice,hohenstein2018architectural,indrasiri2018developing,jacobson2011apis,thielens2013apis,weir2015oracle}    \\ \hline
User Auditing & User auditing can help the API administrator review historical information to analyze who accesses an API, when it is accessed, how it is used, and how many calls are made from the various consumers of the API.~\cite{de2017api} &~\cite{de2017api,gadge2018microservice}    \\ \hline
Username \& Password Authentication & A username and password is the most common form of authentication and is useful when dealing with sensitive data in an API call. In this form of authentication, the client presents the server with a unique name (username) and a secret code (password). The server validates the username and password against its credential store and provides access to the client only on successful validation.~\cite{de2017api} &~\cite{de2017api}    \\ \hline
Web-Based Support & \textit{Type of support where API providers provide support to their consumers, using a website as a communication medium, for example through a live chat functionality.} &~\cite{jacobson2011apis,raivio2011towards}    \\ \hline
X.509 Client Certificate Authentication & An X.509 certificate contains a public key that validates an end entity, such as a web server or an application. It is a good alternative to a username/password for authentication purposes in application-to-application communication. The X.509 certificate contains the identity of the subject.~\cite{de2017api} &~\cite{de2017api}    \\ \hline
XML Threat Protection Policies & XML threat protection policies can be used to check the message payload for the 
following, and reject the message if any of the allowed limits are exceeded.~\cite{de2017api} &~\cite{de2017api}    \\ \hline
\end{longtable}